# Spinodal Decomposition in the $TiO_2$-$VO_2$ System

Zenji Hiroi*, Hiroaki Hayamizu, Toru Yoshida, Yuji Muraoka[†], Yoshihiko Okamoto, Jun-ichi Yamaura[§], and Yutaka Ueda

Institute for Solid State Physics, University of Tokyo, Kashiwa, Chiba 277-8581, Japan

ABSTRACT:
Spinodal decomposition is a ubiquitous phenomenon leading to phase separation from a uniform solution. We show that a spinodal decomposition occurs in a unique combination of two rutile compounds of $TiO_2$ and $VO_2$, which are chemically and physically distinguished from each other: $TiO_2$ is a wide-gap insulator with photo catalytic activities and $VO_2$ is assumed to be a strongly correlated electron system which exhibits a dramatic metal-insulator transition at 342 K. The spinodal decomposition takes place below 830 K at a critical composition of 34 mol% Ti, generates a unidirectional composition modulation along the $c$ axis with a wavelength of approximately 6 nm, and finally results in the formation of self-assembled lamella structures made up of Ti-rich and V-rich layers stacked alternately with 30-50 nm wavelengths. A metal-insulator transition is not observed in quenched solid solutions with intermediate compositions but emerges in the thin V-rich layers as the result of phase separation. Interestingly, the metal-insulator transition remains as sharp as in pure $VO_2$ even in such thin layers and takes place at significantly reduced temperatures of 310-340 K, which is probably due to a large misfit strain induced by lattice matching at the coherent interface.

INTRODUCTION

When two solid compounds that are chemically similar to each other are mixed and heated to high temperatures, they often mix with each other at the atomic scale for any composition. This uniform mixture called the solid solution (SS) can be stable upon cooling to room temperature. On the other hand, in the case that two compounds are rather distinct from each other, they do not mix even at high temperatures, though entropy favors a random distribution of constituents. In an intermediate case, an SS decomposes into two phases upon cooling, the process of which is called the phase separation (PS).

There are two routes to PS.[1] One is the nucleation and growth process, in which tiny aggregates of one of phases begin to form in a matrix of the other phase, finally resulting in a random mixture of the two phases. The other is the spinodal decomposition (SD), in which an SS becomes thermodynamically unstable against a minimal composition fluctuation, and a nearly sinusoidal composition modulation occurs and develops upon cooling or with time duration after quenching from high temperatures.[2] Since the SD takes place "uniformly" with a long-range spatial correlation, quasi-periodic, self-organized two-phase mixtures with various kinds of morphologies are obtained at the nanometer scale. The SD is a general phenomenon observed in various systems such as alloys, polymers, glasses and oxides. It is also believed that the early universe experienced an SD during inflation after Big Bang.[3] Understanding the microscopic mechanism of this ubiquitous event is therefore of widespread interest.

Scientifically interesting and technologically important is resulting microstructures via SD that contain lamellas or bubbles at the nanometer scale. They play a crucial role in determining the chemical and physical properties of composites. In fact, the mechanical properties of some alloys and polymer blends are significantly affected by the morphology.[4] Recently, SD-related microstructures inside nano particles have drawn much attention. For example, the Au-Pt miscibility gap was reevaluated to understand the electrocatalytic properties of core (Au)-shell (Pt) nano particles.[5]

The $TiO_2$-$SnO_2$ system is an SD system most extensively studied among oxides and was only one system known in the large family of rutile compounds. It was first noted by Padurow[6] and studied in detail by Schultz and Stubican.[7] Both the component oxides crystallize in the tetragonal rutile structure (Fig. 1) and form a complete SS above 1723 K. SD takes place below this temperature, giving a miscibility gap with a critical composition of 47 mol% Ti in the phase diagram.[7-9] Resulting modulated structures consist of Ti-rich and Sn-rich phases with composition modulations of wavelengths $\lambda$ = 10-100 nm. Recently, the dielectric and gas-sensing properties of spinodally decomposed (Ti,Sn)$O_2$ thin films were investigated.[10, 11] A unique feature of this spinodal system is the strong anisotropy:[12, 13] a composition modulation occurs selectively along the $c$ axis, reflecting the tetragonal symmetry as well as a smaller lattice mismatch along the $a$ axis (1.5%) than the $c$ axis (3.7%); the modulation direction should be chosen so as to minimize the strain energy at the interface between two phases. This is a great advantage to obtain a well-ordered superlattice by SD. Note that most SDs in cubic alloys or polymer blends possess no such preference in direction, resulting in isotropic mixtures of two phases.

In the present study, we show that the $TiO_2$-$VO_2$ system is another example of SD in the rutile family. The SD of the $TiO_2$-$VO_2$ system was discovered by Zanma and Ueda in 1998, but no report was presented. We are interested in this system because of the distinct differences in their chemical and physical properties: $TiO_2$ without $d$ electrons is a wide band-gap insulator and shows a photocatalytic activity,[14-16] while $VO_2$ with one $d$ electron is a metal at high temperature and becomes an insulator at low temperature; a metal-insulator (MI) transition takes place at $T_{MI}$ = 342 K upon heating.[17] Combining the two compounds in the thin film form has been examined to improve the photoresponse of $TiO_2$.[18] $VO_2$ is potentially important for applications as a ultrafast optical switching device,[19-22] because optical absorption changes across the transition, or as a sensor utilizing the large temperature coefficient of resistivity at the transition.[23] These large differences in properties between the two constituents make the system exceptional; in other SD systems two constituents possess quite similar properties in all aspects. Therefore, it is intriguing to investigate the properties of spinodally decomposed composites in the $TiO_2$-$VO_2$ system.

One important issue to be focused in the present study is concerned with the mechanism of the MI transition of $VO_2$. A large electrical discontinuity of more than several orders of magnitude is observed across the transition.[17, 24, 25] The MI transition is of the first order, accompanied by a structural transition from the high-temperature tetragonal ($P4_2/mnm$) rutile (R) structure to the low-temperature monoclinic ($P2_1/c$) structure (Fig. 1). A striking feature of this monoclinic phase (M1) is the presence of V-V pairs in the strands of edge-sharing octahedra along the $c_r$ axis of the tetragonal structure: alternate V-V separations are 265 and 312 pm in place of the regular 287 pm spacing in the tetragonal structure near the transition.[26] Because this dimerization of V atoms causes a pairing of $3d$ electrons to form a singlet state, the magnetic property of $VO_2$ changes from high-temperature paramagnetic to low-temperature nonmagnetic, so that there is also a large discontinuity in magnetic susceptibility across the transition. On the basis of a large accumulation of experimental and theoretical studies, the mechanism of the MI transition has been greatly debated for decades from various perspectives: interpretations based on Peierls,[26, 27] Mott-Hubbard[28, 29] and other mechanisms[30] have been given, depending on how much one considers the importance of electron correlations. Although a general consensus may have not yet been reached, one can definitely say that the structural transition plays a crucial role in the mechanism of the MI transition of $VO_2$.

A number of workers have investigated the influence of cation substitutions on $T_{MI}$ and the crystal structures.[26] For example, the Cr substitution up to 2.5 mol% makes the $T_{MI}$ slightly higher by ~5 K and introduces another insulating monoclinic phase (M2) between the R and M1 phases;[28, 31] the MI transition occurs between the R and M2 phases in the Cr-substituted compounds. In the M2 phase, V atoms are split into two sets of chains parallel to the $c_r$ axis. On one set of chains V atoms are paired with alternating long and short separations, as in those of the M1 phase, while, on the other set of chains, the V-V atoms form a zig-zag pattern but remain equally spaced.[31] Thus, the M2 structure may be considered as an intermediate structure between the R and M1 structures. The substitution of most elements for V seems to be limited below 20 mol%, above which either other phases are stabilized or conventional two-phase mixtures are obtained. On the Ti substitution, however, it was reported that complete SSs $V_{1-x}Ti_xO_2$ could be prepared in polycrystalline samples or thin films;[32, 33] $Ti_xV_{1-x}O_2$ may be a right chemical formula according to the rules provided by the International Union of Pure and Applied Chemistry (IUPAC),[34] but we use here $V_{1-x}Ti_xO_2$ following previous studies. In a later study on the single crystal growth, however, the presence of a complete SS was questioned because single crystals were obtained only for $x > 0.8$ even from nominal compositions of $x > 0.5$.[35] On the other hand, in the V-rich side, $V_{1-x}Ti_xO_2$ with $x \leq 0.06$ was prepared, where an $x$-$T$ phase diagram including the R, M1 and M2 phases is obtained,[36] as in the case of $V_{1-x}Cr_xO_2$. Therefore, both the Ti- and V-rich sides of the phase diagram are well understood, but the middle region remains less explored.

We show clearly that there is a miscibility gap in the $TiO_2$-$VO_2$ phase diagram: an SD occurs in a similar manner as in the $TiO_2$-$SnO_2$ system but below a much lower temperature of 830 K. MI transitions are not observed in SSs with $0.3 < x < 0.6$ obtained by rapid quenching from high temperature but appear after annealing at lower temperatures, where spinodally modulated structures with $\lambda \sim 6$ nm or phase-separated lamellar structures with $\lambda = 30$-$50$ nm are observed. The MI transition occurs within the resulting thin V-rich layers at reduced temperatures of 310-340 K. Such large reductions of $T_{MI}$ are quite unusual, which may be due to a large anisotropic strain caused by lattice matching at the interface between the V-rich and Ti-rich layers. A linear relationship between the $T_{MI}$ and tetragonality is established, which may give a hint for understanding the mechanism of the MI transition of $VO_2$.

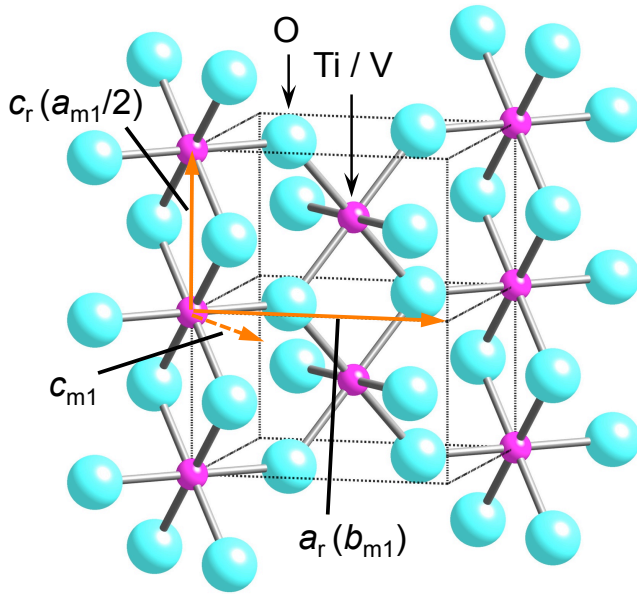

Figure 1. Rutile-type crystal structures of $TiO_2$ and $VO_2$ in the metal phase. Strands of edge-sharing Ti/V-O octahedra are aligned along the $c_r$ axis of the tetragonal structure. The relation between the unit cells of the tetragonal rutile and the monoclinic M1 phases of $VO_2$ is shown.

EXPERIMENTS

Polycrystalline samples of $V_{1-x}Ti_xO_2$ were prepared from $TiO_2$, $V_2O_5$ and $V_2O_3$, the last of which had been obtained by reducing $V_2O_5$ in a hydrogen flow at 873 K. The three materials were mixed in an appropriate ratio in an agate mortar, pressed into a pellet, sealed in a quartz ampoule, and heated at 873 K for 1 day. The obtained pellet was ground, pressed again into a pellet, and annealed in a quartz ampoule at 1173 K for 2 days. A solid solution was obtained by rapidly quenching the ampoule into ice water; slower cooling resulted in a partially decomposed sample. Phase separation was induced by annealing the quenched pellets at low temperatures: the pellet was sealed in a quartz ampoule, heated at 1173 K for 1 h and then annealed at 473, 573, 673 and 773 K for 1-12 h, followed by a rapid quench into ice water.

A single crystal was also grown by the floating-zone method. A polycrystalline rod of a solid solution with $x$(ss) = 0.4 was partially melted and slowly solidified in a rate of 1 mm/h in an Ar flow. The obtained boule was annealed at 1323 K and quenched into ice water to obtain a uniform solid solution. Decomposition was performed at 673 K for 12 h.

All products were examined by means of powder X-ray diffraction (XRD) using Cu-K$\alpha$ radiation. Transmission electron microscopy observations were carried out in a JEOL-2010F electron microscope equipped with an energy-dispersive X-ray (EDX) analyzer. A specimen for this was prepared by ion-milling a fragment of the single crystal. To observe the MI transition, magnetic susceptibility was measured at a magnetic field of 1 T in a Quantum-Design magnetic property measurement system, and resistivity was measured in a Quantum-Design physical property measurement system.

SOLID SOLUTIONS

Phases in quenched samples and their average crystal structures are examined by means of powder XRD at room temperature. As shown for selected Ti contents in Fig. 2, all quenched samples are mono phasic without a trace of impurities or a signature for phase separation, implying that complete SSs exist at high temperature and can be quenched to room temperature. The samples with $x \leq 0.15$ takes the monoclinic M1-type structure, which is evidenced by the presence of the (-102) reflection at $2\theta \sim 33.5°$. On the other hand, the $x = 0.20$ sample takes the M2-type structure, and all the other Ti-rich samples take the tetragonal rutile structure. The lattice parameters are given in Table 1.

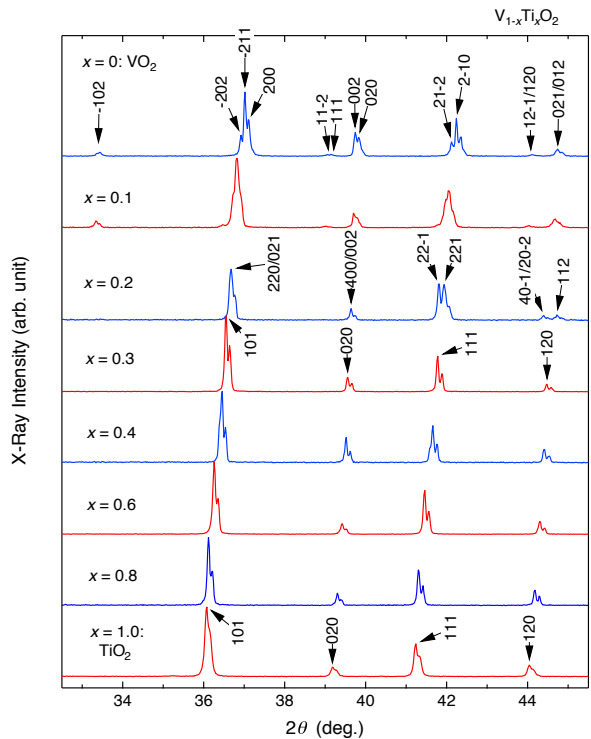

Figure 2. Powder X-ray diffraction patterns measured at room temperature of VO$_2$, TiO$_2$ and their SSs V$_{1-x}$Ti$_x$O$_2$ obtained by rapid quenching from 1173 K. Peak indices are given on the bases of the M1 structure for $x$ = 0 and 0.1, the M2 structure for $x$ = 0.2, and the rutile structure for $x$ = 0.3-1.

The composition dependences of the lattice parameters, which are determined by the least square method using the peak positions, are plotted in Fig. 3; for the monoclinic structures, those of the corresponding tetragonal cells are given: $a_r = (b_{m1} + c_{m1}\sin\beta)/2$ and $c_r = a_{m1}/2$ for the M1 phase, and $a_r = (a_{m2}/2 + c_{m2}\sin\beta)/2$ and $c_r = b_{m2}/2$ for the M2 phase.[31] The $a$-axis length of the SSs shows a nearly linear $x$ dependence, while the $c$-axis length tends to saturate toward $x = 1$. This deviation from the conventional Vegard's law for the $c$-axis length may indicate a difference in chemical bonding along the $c$ axis between the end compounds. The unit-cell volume increases gradually with increasing $x$, which approximately follows a quadratic relation $V$ ($10^{-3}$ nm$^3$) = 59.02 + 5.0$x$ − 1.6$x^2$. Later we will utilize this relation to estimate local Ti contents in phase-separated samples.

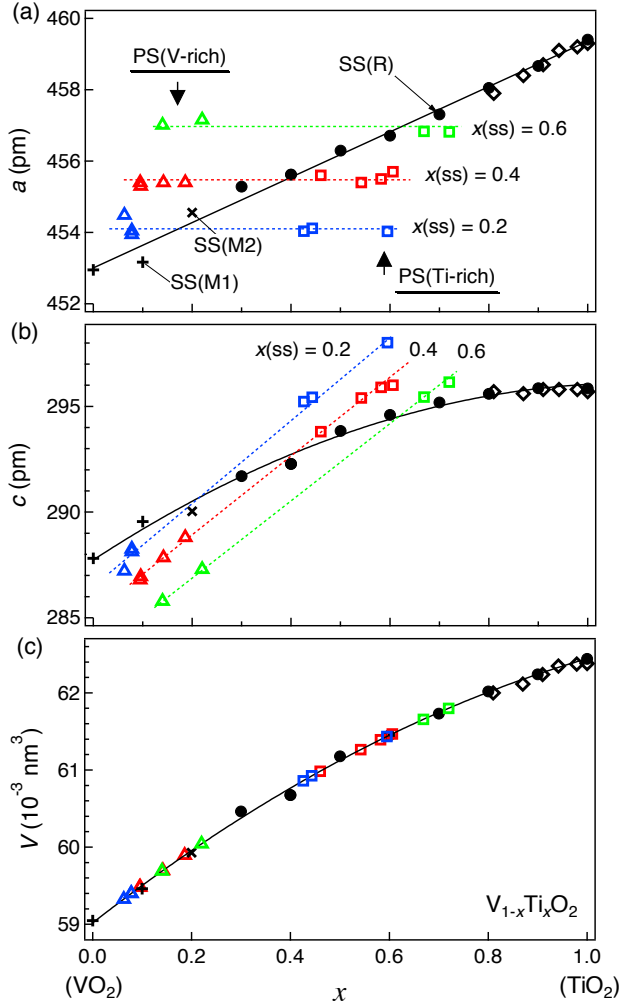

Figure 3. Composition dependences of the tetragonal lattice parameters. Data from literature for Ti-rich compositions are also plotted by open diamonds.[35] The solid line for $a$ is a linear fit: $a$/pm = 453.0(1) + 6.4(2)$x$. Those for $c$ and $V$ are fits to quadratic forms: $c$/pm = 287.7(2) + 15.3(1)$x$ − 7(1)$x^2$ and $V$/$10^{-3}$ nm$^3$ = 59.02(4) + 5.0(2)$x$ − 1.6(2)$x^2$. Also plotted by open marks are the lattice constants of the V-rich (triangles) and Ti-rich (squares) phases in PS samples obtained by annealing SSs with $x$(ss) = 0.2, 0.4 and 0.6. The Ti contents of the two phases have been decided so as to reproduce the cell volume of the SS in (c). The dotted lines in (a) and (b) are linear guides to the eyes.

Phase transitions in (V, Ti)O$_2$ are probed by magnetic susceptibility $\chi$. Pure VO$_2$ shows a sudden and large jump in $\chi$ at 342 K upon heating, as shown in Fig. 4, where insulator-to-metal and structural transitions from M1 to R take place simultaneously. Here the MI transition temperature $T_{MI}$ is defined as a midpoint of the jump in $\chi$ upon heating; there is a thermal hysteresis, reflecting the first-order character, and a drop in $\chi$ occurs at a slightly lower temperature of 338 K upon cooling. The low-temperature $\chi$ of VO$_2$ is nearly equal to that of TiO$_2$ [Fig. 4(b)], because all V spins form pairs to become nonmagnetic in total as a result of the structural dimerization in the M1 phase. In contrast, Ti-substituted samples such as $x$ = 0.05 and 0.10 exhibit two-step transitions, those at low and high temperatures corresponding to the M1-M2 and M2-R phase transitions, respectively; only half of V(Ti) chains are dimerized in the M2 phase, so that the M2 phase has an intermediate $\chi$ value. As $x$ increases further, the jump becomes obscure, and no changes are discernible for $x \geq 0.30$. On one hand, large increases in $\chi$ toward $T = 0$ are observed for $0 < x < 1$ [Fig. 4(a)], which are approximately proportional to $1/T$ and are ascribed to nearly free V spins that have lost their partners by the Ti substitution, as observed in $V_{1-x}Nb_xO_2$.[37] Corresponding to the $\chi$ data, as shown in Fig. 5, resistivity shows a sudden drop at $T_{MI}$ either for pure $VO_2$ or the $x = 0.1$ sample but just semiconducting behavior for $x \geq 0.2$. Note that the drop in $\rho$ of the $x = 0.1$ sample occurs at the higher-temperature jump in $\chi$ for the M2-R transition.

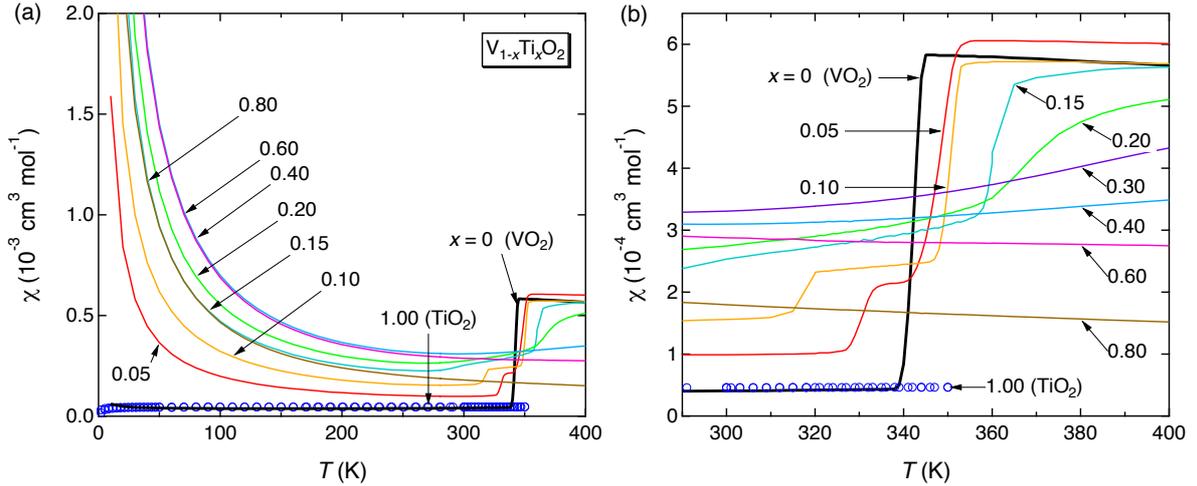

Figure 4. Magnetic susceptibility $\chi$ of selected SSs in a wide temperature range (a) and near the MI transition (b). All the measurements were carried out upon heating in a magnetic field of 1 T. A sudden jump is observed for pure $VO_2$ at 342 K, and two-step transitions are clearly detected for $x = 0.05$ and 0.10: the low- and high-temperature jumps correspond to the M1-M2 and M2-R transitions, respectively, with the MI transition occurring at the latter (Fig. 5). As $x$ increases further, the jump becomes obscure and completely disappears for $x \geq 0.30$. Large Curie tails are observed at low temperature toward $T = 0$ only for SSs, which come from orphan V spins with Ti neighbors in the structurally paired sites of the M1 structure.

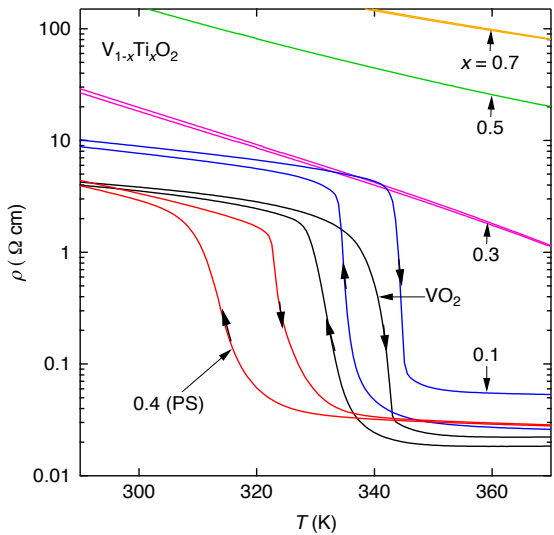

Figure 5. Resistivity $\rho$ measured on polycrystalline samples of pure $VO_2$ and four SSs with $x = 0.1$, 0.3, 0.5 and 0.7. Measurements were performed upon cooling from 400 to 4 K and then heating to 400

K, as indicated by arrows on each data curve. The $\rho$ of a PS sample obtained by annealing the $x$(ss) = 0.4 sample at 673 K for 12 h is also shown.

A $T$-$x$ phase diagram is obtained from the $\chi$ data (Fig. 6), which is consistent with previous ones:[36] the M1-M2 boundary goes down and the M2-R boundary goes up with increasing $x$. Note that the MI transition takes place at the latter. The Ti substitution may stabilize the insulating state and also makes the M2 phase more stable instead of the M1 phase, as in the case of the Cr substitution.[31] For $x > 0.2$ $T_{MI}$ seems to increase further, but the phase relation becomes unclear in these SSs, as reported in previous study.[32] This must be related to the presence of the miscibility gap mentioned in the next section: there must be a tendency towards PS even in quenched samples at compositions within the miscibility gap.

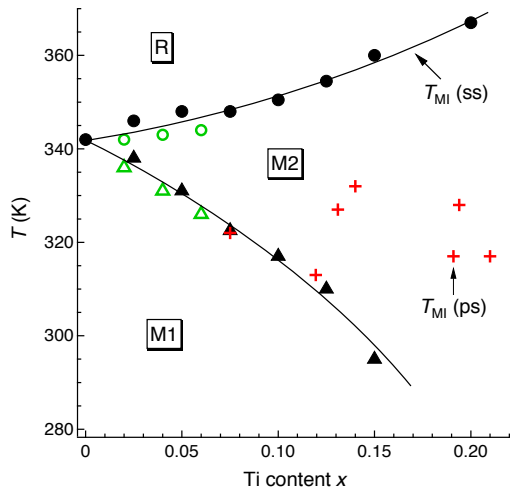

Figure 6. $T$-$x$ phase diagram for $V_{1-x}Ti_xO_2$ SSs determined by the magnetic susceptibility data of Fig. 4. The open marks are from literature.[36] The M2 phase becomes more stable with increasing $x$ between the R and M1 phases. The crosses indicate the $T_{MI}$s of PS samples plotted against the Ti content of the V-rich phase.

SPINODAL DECOMPOSITION

To investigate the thermal stability of quenched SSs with $x$(ss) = 0.2, 0.3, 0.4, 0.5 and 0.6, heat treatments were carried out at $T_a$ = 848, 773, 673, 573 and 473 K. Changes in the XRD pattern with duration were not observed at 848 K, but at the lower temperatures, as typically shown in Fig. 7 for $x$(ss) = 0.4 and $T_a$ = 673 K. After annealing for 1 h, sideband peaks called satellites appear on either side of certain fundamental reflections such as (101), (111) and (121), which evidences the formation of an additional compositional and/or structural modulation with a long wavelength $\lambda$. Since the satellites are observed only for $hkl$ reflections with $l \neq 0$, the modulation must occur along the $c$ axis, as observed in the $TiO_2$-$SnO_2$ system.[12] Provided this, the $\lambda$ is calculated to be ~6 nm from the positions of the satellite peaks beside the (101) peak: $(1/d_+)^2 + (1/d_-)^2 = 2[(1/d_{101})^2 + (1/\lambda)^2]$, where $d_+$, $d_-$ and $d_{101}$ are the interplanar spacings of the high-angle satellite, the low-angle satellite, and the (101) peak, respectively.[9] The selection of the $c$ axis is reasonable, because the lattice mismatch between $TiO_2$ and $VO_2$ is smaller along the $a$ axis (0.86%) than the $c$ axis (3.6%).

Further annealing at 673 K for 12 h results in the lack of original peaks and their satellites, and alternatively splitting into two sets of peaks, indicating that a complete phase separation has been induced, as schematically illustrated in Fig. 7(b). The two sets of peaks at low- and high-angle sides are indexed to a tetragonal structure and a monoclinic structure of the M1 type, respectively. Obviously, they correspond to the Ti-rich and V-rich phases in the decomposed sample, respectively. Note that the (020) and (220) peaks are not split, which means that the $a$-axis lengths remain equal between the two phases. This is definitely due to lattice matching at the coherent interface. Since the $a$-axis length is larger for $TiO_2$ than $VO_2$, this lattice matching should cause a misfit strain that forces the $c$-axis length to increase and decrease in the Ti-rich and V-rich phases, respectively, so as to keep the cell volume unchanged. The lattice parameters of each phase are given in Table 1. No more change was observed by extended annealing in the present $TiO_2$-$VO_2$ system, which is not the case for the $TiO_2$-$SnO_2$ system where an incoherent structure is finally attained by introducing misfit dislocations.[9] Thus phase-separated samples are completely transformed to uniform SSs by annealing at higher temperatures above 848 K; the SD occurs reversibly.

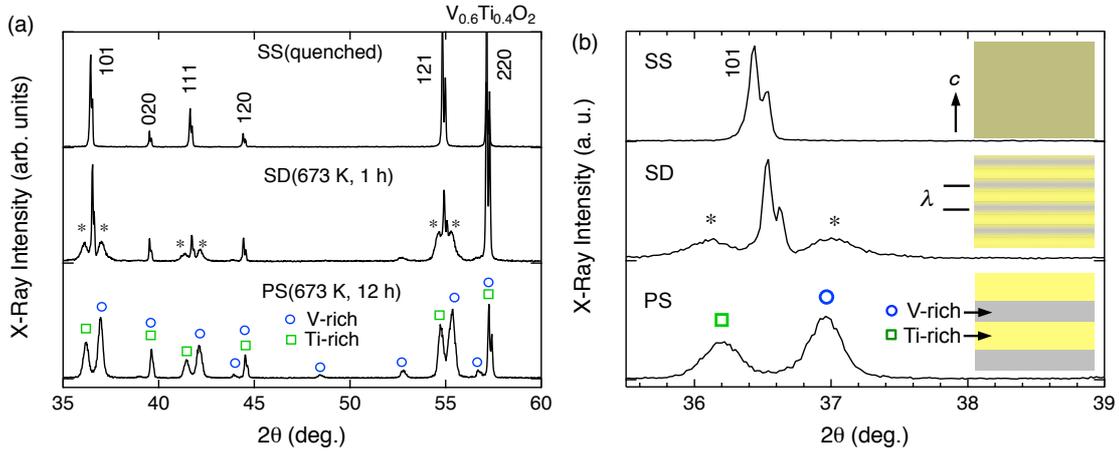

Figure 7. (a) Evolution of XRD patterns of $V_{0.6}Ti_{0.4}O_2$ measured at room temperature with the duration of annealing at 673 K. Starting from the quenched SS (SS, top), a spinodal decomposition takes place after 1 h (SD, middle), which is evidenced by the appearance of satellite peaks marked by * on both sides of some fundamental reflections, and then a complete phase separation occurs after 12 h (PS, bottom), where two sets of diffraction peaks coexist: one is from a Ti-rich phase with the R structure (squares) and the other is from a V-rich phase with the M1 structure (circles). The XRD profiles around the (101) reflection are expanded in (b). The evolution of microstructures with duration is schematically depicted in the inset: a composition modulation occurs along the $c$ axis with a wavelength $\lambda \sim 6$ nm and finally results in an alternate stacking of Ti-rich and V-rich layers.

A microscopic phase separation after annealing for long time has been confirmed in electron microscopy. As clearly shown in the high-resolution image of Fig. 8 for an $x(ss) = 0.4$ crystal annealed at 673 K for 12 h, bright and dark bands, each of a few nm thick, alternate quasi-periodically along the $c$ axis. The interface is definitely perpendicular to the $c$ axis, because it appears as a sharp line when incident electrons run exactly parallel to the [110] zone axis. The stacking sequence is not completely periodic but a few "dislocations" are observed in Fig. 8(a). Moreover, the thickness fluctuates from place to place, and the periodicity is approximately 30-50 nm. Note that the bright bands are always thicker than the dark bands.

Chemical analyses for Ti and V were carried out by EDX using a normal electron probe of ~1000 nm φ and a microprobe of ~1 nm φ on a selected area shown in Fig. 8(b). The former gave an average Ti/V ratio approximately 2/3 as intended. In the latter microprobe analyses shown in Figs. 8(c) and (d), it is clearly noticed that Ti condenses in thick bright bands and V does in thin dark bands. The composition profiles along the $c$ axis of Fig. 8(e) reveal anti-phase modulations for Ti and V contents. They show gradual variations across the interface, but this may be due to a limited spatial resolution mostly due to the finite size of electron probes and their spread inside the crystal. Since the high-resolution image shows a sudden change in contrast across the interface, the composition profile may not be sinusoidal but close to the square-well type. Therefore, the phase-separated sample consists of Ti-rich and V-rich lamellas stacked alternately along the $c$ axis, as schematically depicted in the inset of Fig. 7(b). This microstructure naturally originates from spinodal decomposition at intermediate annealing. Note that the wavelength is increased from 6 nm in the spinodal regime to 30-50 nm in the final phase separation. The wavelength in the PS state is similar to those observed in $(Ti,Sn)O_2$.[10, 11]

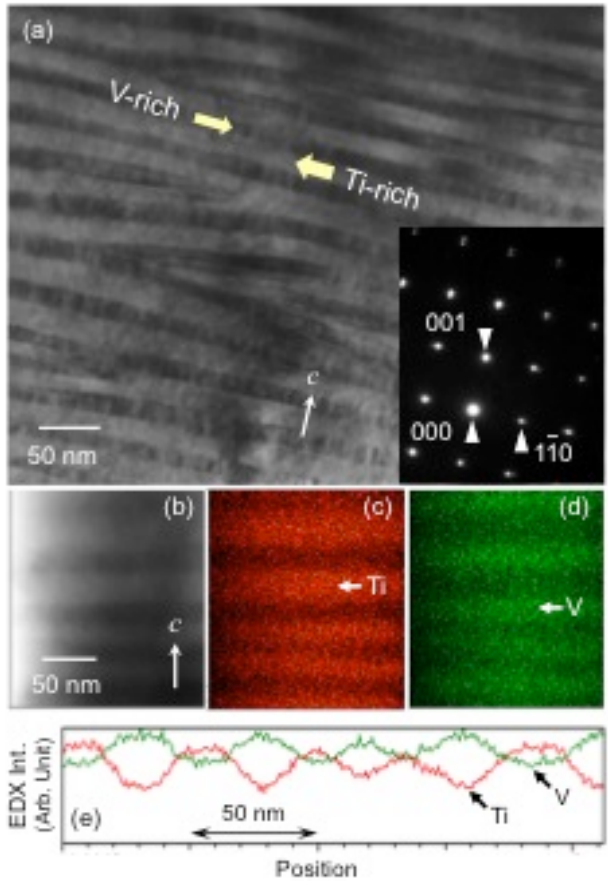

Figure 8. Electron micrograph taken with the [110] zone axis for a $V_{0.6}Ti_{0.4}O_2$ crystal after annealing at 673 K for 12 h. The corresponding electron diffraction pattern is shown in the inset of (a). Elemental mapping by using a microprobe in EDX was performed on a selected area (b) for Ti (c) and V (d), where a brighter region contains more Ti/V content. Line scans of the EDX intensity along the $c$ axis in (e) show alternating variations of Ti and V contents with a wavelength of approximately 45 nm.

It is rather difficult to decide the composition quantitatively in the two domains in PS samples by using EDX. We do this by reference to the unit cell volume of quenched SSs. Provided that the cell volume is an invariable at a certain composition in a uniform mixture and gives a measure for the composition, one can estimate the composition of a domain with reference to the $x$ dependence of the cell volume determined for uniform SSs shown in Fig. 3(c). For example, the PS sample of $x$(ss) = 0.4 annealed at 673 K is composed of a Ti-rich phase with $V = 0.061263$ nm$^3$ and a V-rich phase with $V = 0.059697$ nm$^3$ (Table 1). From these $V$ values, one can estimate the Ti content $x$(ps) of each phase by using the quadratic relation for $V$ shown in Fig. 3(c): $x$(ps) = 0.54 and 0.14 for the Ti-rich and V-rich phases, respectively. In addition, the $x$(ps) varies as 0.46-0.61 and 0.10-0.19, respectively, depending on the annealing time. All the $x$(ps)s thus decided for various $x$(ss)s and annealing conditions are given in Table 1 and are also plotted in the phase diagram of Fig. 9. The lattice parameters of V-rich and Ti-rich phases in PS samples are plotted in Fig. 3 against the Ti content. As shown in Fig. 3(a), the $a$(ps) either of $x$(ss) = 0.2, 0.4 or 0.6 remains nearly constant over a wide variation of $x$ at almost the same value as the original SS value. In contrast, the $c$(ps) is significantly decreased in the V-rich phase and increased in the Ti-rich phase in comparison with the $c$(ss) at the same Ti content. This is obviously due to misfit strain at the coherent interface mentioned above.

The subsolidus miscibility gap of the $TiO_2$-$VO_2$ system is shown in Fig. 9. The miscibility dome determined by the annealing experiments on the $x$(ss) = 0.4 sample is nearly symmetric. The critical composition is 34 mol% Ti, and the critical temperature is 830 K; for comparison, they are 47 mol% Ti and 1703 K in the TiO$_2$-SnO$_2$ system.[8] The regular-solution model gives a coherent spinodal temperature as $T_s = C(1 - x)x$, where $C$ is a constant that depends on elastic constants, linear expansion coefficients, etc. The experimental spinodal dome of the TiO$_2$-SnO$_2$ system is reproduced by $T_s = 6392.5(1 – x)x$ after shifting 0.3 mol% toward the SnO$_2$ side.[8] However, this relation applies only at high temperatures above 1473 K because of sluggish solid-state kinetics at low temperature. The spinodal dome of the TiO$_2$-VO$_2$ system can be fitted to the form $T_s = 3330(1 – x)x$ with a 16 mol% shift toward the VO$_2$ side. However, the experimental curve shows a significant deviation from the calculated one even near the critical temperature, reflecting slow kinetics at these low temperatures. The observed large shift of the critical composition from 50 mol% may indicate a large difference in elastic properties between the end members. Separated compositions after SD starting from other SSs with $x$(ss) = 0.2 0.3 0.5 and 0.6 are also plotted in Fig. 9, which mostly follow the above spinodal line; some data points inside the dome may be due to the slow kinetics, and some outside of the dome, particularly in the Ti-rich side, may come from ambiguity in determining the local Ti content from the $x$ dependence of the cell volume which tends to saturate toward TiO$_2$.

It would be helpful now to describe a typical microstructure for a PS sample. As depicted in the inset of Fig. 9 for the $x$(ss) = 0.4 SS annealed at 673 K for 12 h, the compositions of the V-rich and Ti-rich phases are V$_{0.86}$Ti$_{0.14}$O$_2$ and V$_{0.44}$Ti$_{0.56}$O$_2$, respectively, and the wavelength is 30-50 nm. Since the average Ti content should be 0.4, the volume fraction between them becomes 7:13, approximately 1:2. This ratio is close to the thickness ratio between the two kinds of lamellas in the electron micrograph of Fig. 8: the narrow dark and wide bright bands correspond to V$_{0.86}$Ti$_{0.14}$O$_2$ and V$_{0.46}$Ti$_{0.56}$O$_2$, respectively. Provided $\lambda$ = 40 nm, their average thicknesses turns out to be 14 and 26 nm.

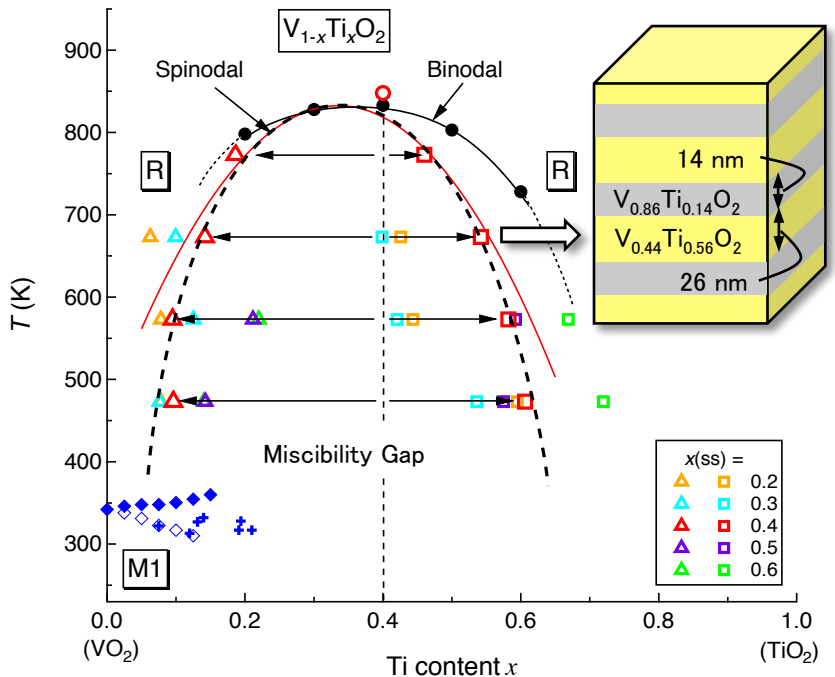

Figure 9. Phase diagram of the TiO$_2$-VO$_2$ system. A binodal line decided by Zamma and Ueda and a coherent spinodal line determined on the basis of annealing experiments on the $x$(ss) = 0.4 SS at four temperatures are shown: triangles and squares represent the compositions of the V-rich and Ti-rich phases in the decomposed samples, respectively. The results of other annealing experiments on SSs with $x$(ss) = 0.2, 0.3, 0.5 and 0.6 are also shown. The solid red line represents a fit to the regular-solution model: $T_s = 3330[1 – (x + 0.16)](x + 0.16)$. The critical temperatures of the R-M2 (solid diamond) and M2-M1 (open diamond) transitions observed in SSs are shown in the left-bottom,

together with those of the MI transitions (cross) observed in the V-rich phase in the PS samples. A microscopic structure of the PS sample annealed at 673 K for 12 h is schematically depicted in the inset.

METAL-INSULATOR TRANSITION

Next, we address how the MI transition is observed as the spinodal decomposition proceeds with duration. The evolution of magnetic susceptibility for a series of samples with $x$(ss) = 0.4 is shown in Fig. 10. A jump of $\chi$ associated with the transition is not present in the SS and emerges at 332 K for the SD sample after annealing at 673 K for 1 h. A nearly same jump is observed at a slightly higher temperature of 333 K in the PS sample annealed further for 12 h. This means that the MI transition is induced as the result of phase separations through which many of V atoms mixed with Ti atoms have condensed into the V-rich phase. In fact, one expects an MI transition in the V-rich phase including 14 mol% Ti as SSs of $x < 0.2$ exhibit MI transitions (Fig. 6). Note that the resistivity of the PS sample shows a sudden drop at approximately the temperature of the jump in $\chi$ (Fig. 5).

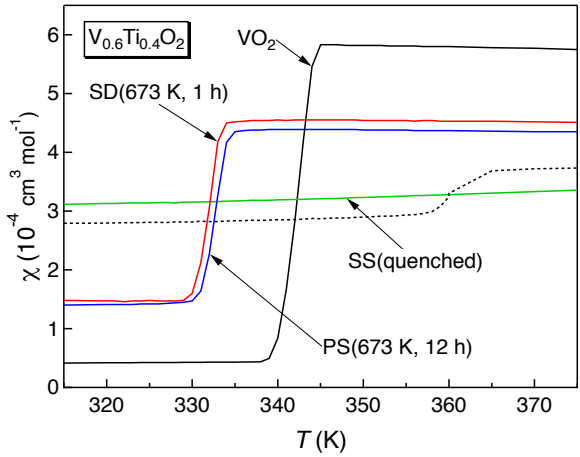

Figure 10. Evolution of the magnetic susceptibility of $V_{0.6}Ti_{0.4}O_2$ with the duration of annealing. Sudden jumps are observed for the SD and PS samples at 332 and 333 K, respectively, in contrast to the flat curve of the corresponding SS. The dotted curve represents a simulated $\chi$ that is a sum of two components from a V-rich phase with $x = 0.15$ and a Ti-rich phase with $x = 0.60$ in a molar ratio of 7:13 on the basis of the $\chi$ data of SSs.

The MI transition in the PS sample is accompanied by a structural transition. The powder XRD profile of Fig. 7 measured at room temperature reveals that the V-rich phase has the M1 structure, while the Ti-rich one takes the rutile structure. We confirmed by means of XRD at elevated temperatures that the V-rich phase exhibited a transition to the rutile structure above the $T_{MI}$. However, we did not detect the M2 phase between the M1 and R phases, as in the case of SSs. Correspondingly, the $\chi$ shows a single-step transition (Fig. 10), in spite that the corresponding SS with $x$(ss) = 0.15 shows a two-step transition (Fig. 4). Thus, it is likely that the M2 phase is suppressed owing to lattice matching with the adjacent tetragonal Ti-rich layers. This may be because the M2 structure requires a non-orthogonal axis at the interface, i.e. 90.5°,[31] whereas the M1 structure can share a common interface of tetragonal symmetry with the rutile structure, as depicted in Fig. 1.

It is rather surprising that such a sharp transition as in Fig. 10, as sharp as in pure $VO_2$, is observed in the thin layers of the V-rich phase: the thickness is approximately 14 nm in the PS sample and even less than 6 nm in the SD sample. Particularly, in the latter case, the very thin V-rich layer is not completely isolated from the Ti-rich layer but may exit in a sinusoidal composition modulation. This strongly suggests a local character of the transition. However, there are two important differences in the transition between pure $VO_2$/SS and SD/PS samples. One is on the transition temperature and the other is on the magnitude of jump in $\chi$.

The $T_{MI}$ of the V-rich phase in the PS sample is reduced by 9 K from that of pure $VO_2$ and even by 27 K from that of the corresponding SS with $x$(ss) = 0.15. These large reductions in $T_{MI}$ must be ascribed to anisotropic strain caused by lattice matching at the interface. Previous studies on $VO_2$ revealed that isotropic pressure enhances $T_{MI}$ by 0.6 K/GPa, whereas anisotropic compression along the $c$ axis is more effective to reduce $T_{MI}$ by -12 K/GPa.[38] However, an actually observed decrease in $T_{MI}$ was only 0.6 K at 0.05 GPa because of experimental difficulty in applying anisotropic pressure on a crystal. In the present study, the V-rich layer in the PS sample should suffer tensile stress from the adjacent Ti-rich layers, as schematically depicted in Fig. 11, because the in-plane lattice constant is smaller for the V-rich layer. This results in compression along the $c$ axis in the V-rich layer as far as the coherent interface is kept. In fact, as shown in Fig. 3, the $a$-axis lengths coincide between the two phases, and the tetragonality $c/a$ of the V-rich phase is significantly reduced from 0.6354 for pure $VO_2$ down to 0.625, as shown in Fig. 11. The observed reduction of $T_{MI}$ by 27 K may correspond to a huge anisotropic pressure of 2.2 GPa according to the previous result.[38] This value is too large to achieve in conventional experiments, but can be possible in a thin layer by a coherent lattice matching in the PS sample. The slightly lower $T_{MI}$ of the SD sample may indicate a larger strain in the thinner layer of the SD sample.

We have also studied MI transitions in other SSs with $x$(ss) = 0.2, 0.3, 0.5, and 0.6. Compared with the case of $x$(ss) = 0.4, larger decreases in $T_{MI}$ and $c/a$ are observed especially for the $x$(ss) = 0.5 and 0.6 samples, in which the volume fraction of the V-rich phase is relatively small and thus one expects a larger tetragonality. Figure 11 shows the relation between $T_{MI}$ and $c/a$ for the PS samples. The previous results for a pure $VO_2$ film grown on the (001) or (110) surface of the $TiO_2$ substrate are also plotted: $T_{MI}$ is decreased on (001) and increased on (110), as the $c$ axis is compressed and expanded by the epitaxial strain, respectively.[39] Moreover, the small increase of $T_{MI}$ with $x$ in SSs may be due to the increase of $c/a$. All these data indicate that there is a linear relationship between $T_{MI}$ and $c/a$, suggesting that the tetragonality plays a crucial role in the mechanism of the MI transition of $VO_2$.[40] Probably, the electron correlation is not a main player but a certain structural instability inherent to the rutile structure drives the MI transition of $VO_2$, which will be discussed elsewhere.

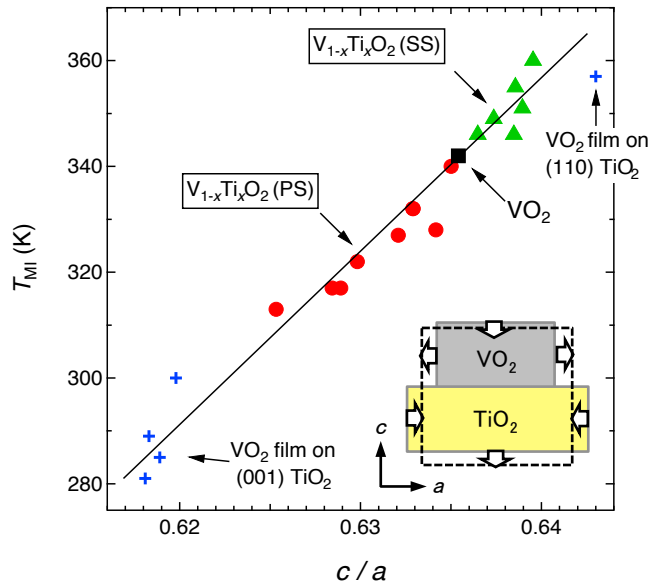

Figure 11. Metal-insulator transition temperature $T_{MI}$ against the tetragonality $c/a$ of the rutile structure. The $T_{MI}$s of pure $VO_2$, solid solutions, phase-separated samples, and $VO_2$ thin films grown on the (001) and (110) $TiO_2$ substrates[39] are plotted. There is a linear relationship between the $T_{MI}$ and $c/a$.

The second important difference in the transition between pure $VO_2$/SS and SD/PS samples is the large jump in $\chi$ for the latter in Fig. 10: it reaches about half of that of pure VO$_2$. This is strange because the volume fraction of the V-rich phase is merely 7/20, as illustrated in Fig. 9. More precisely, as the actual Ti content of the V-rich phase is 0.14, one expects a 15% jump, judging from the $\chi$ data of the corresponding $x$(ss) = 0.15 SS in Fig. 4(b). Also taking account of the $\chi$ of the $x$(ss) = 0.6 SS as representing that of the Ti-rich phase with $x$(ps) = 0.56, one can simulate the $\chi$ curve of the PS sample by summing the two contributions, which is shown by a broken line in Fig. 10. Apparently, the observed jump is more than three times as large as calculated one. The origin of the larger jump must come from two factors: larger $\chi$ above $T_{MI}$ and smaller $\chi$ below $T_{MI}$ compared with the calculated curve. The former may imply enhanced Pauli susceptibility from the metallic V-rich phase, and the latter must be ascribed to smaller contributions from orphan V spins that have failed to make a pair with another V spins in the M1 structure. In fact, the amounts of free V spins are only 9% for the PS sample and 18% for the corresponding SS, which are estimated from Curie tails at low temperature below 100 K. We think that these give evidence of a certain preferential arrangement of Ti and V atoms in the SD/PS structure that are different from a random distribution expected for an SS. Probably, more V-V pairs are generated in the process of SD when a composition modulation grows along the $c$ axis. This may also affect the Pauli susceptibility in the metallic phase. Therefore, an unusual MI transition occurs in the SD/PS samples. For further discussion, we need information from a microscopic probe such as NMR on V nuclei.

CONCLUSION AND FINAL REMARKS

We report on the spinodal decomposition of the binary TiO$_2$-VO$_2$ system. It is a unique SD system, because the end members have very different properties: TiO$_2$ is a wide-gap insulator and VO$_2$ exhibits a dramatic metal-insulator transition at 342 K. A miscibility gap is observed below a critical temperature of 830 K at a critical composition of 34 mol% Ti. The SD occurs unidirectionally along the $c$ axis of the tetragonal rutile structure and proceeds with the duration of annealing to generate a composition modulation with a wavelength typically 6 nm for $x$(ss) = 0.4. Finally, phase-separated, quasi-periodic microstructures composed of Ti-rich and V-rich lamellas are attained; a typical PS sample contains V$_{0.86}$Ti$_{0.14}$O$_2$ layers of 14 nm thick and V$_{0.44}$Ti$_{0.56}$O$_2$ layers of 26 nm thick. An MI transition is not observed in SSs with intermediate compositions but in the thin V-rich layers of the SD/PS samples at significantly reduced transition temperatures compared with those of pure VO$_2$ and the corresponding SSs. The reductions of $T_{MI}$ are ascribed to the enhancement of tetragonality due to a large misfit strain generated at the coherent interface between the two component layers. We find a linear relation between $T_{MI}$ and $c/a$, suggesting a crucial role of the crystal structure on the mechanism of the MI transition of VO$_2$.

We are interested in utilizing these spinodal microstructures for future applications. The control of $T_{MI}$ by using the coherent strain may be promising for an ultrafast optical switching device and a sensor utilizing the large temperature coefficient of resistivity at the transition. The epitaxial strain in thin films gives an alternative route for this, but the advantage of SD is its easy-to-use approach: just annealing SSs yields PS automatically in the bulk form; a mega-layered structure can be reversibly produced in a crystal by a self-organized way. Thermoelectric properties must be also intriguing to be explored. Furthermore, we are searching for an SD system in other combinations of rutile compounds. There are many fascinating compounds in the family: CrO$_2$ is a well-known half-metallic ferromagnet, and RuO$_2$ is a good electric conductor and serves as an efficient catalyst. Novel combinations between two rutile oxides with different properties would open up new possibility for materials science. We would like to establish "spinodal technology" in our future work.

AUTHOR INFORMATION
Corresponding Author
*E-mail: hiroi@issp.u-tokyo.ac.jp
Present Address
†Present address: Graduate School of Natural Science and Technology, Okayama University,

Tsushimanaka, Kita-ku, Okayama 700-8530, Japan
§Present address: Tokyo Institute of Technology, Materials Research Center for Element Strategy, Nagatsuta, Yokohama, Kanagawa 226-8503, Japan

Notes
The authors declare no competing financial interest.

ACKNPWLEDGMENTS
We thank M. Ichihara for his help in electron microscopy observations.
14

Table 1. Crystallographic parameters and the metal-insulator transition temperature $T_{MI}$ of selected solid solutions $V_{1-x}Ti_xO_2$ and phase-separated samples starting from a solid solution with $x(ss) = 0.4$. The lattice parameters are given on the basis of the tetragonal rutile structure ($P4_2/mnm$).

| Sample | $T_a$ (K) | Ti content $x$ | $a$ (pm) | $c$ (pm) | $V$ ($10^{-3}$ nm$^3$) | $c/a$ | $T_{MI}$ (K) |
|---|---|---|---|---|---|---|---|
| $VO_2$ | | $0^c$ | 452.95 | 287.81 | 59.048 | 0.635 | 342 |
| SS$^a$ | | $0.1^d$ | 453.17 | 289.55 | 59.461 | 0.639 | 350 |
| | | $0.2^e$ | 454.56 | 290.04 | 59.928 | 0.638 | 367 |
| | | 0.4 | 455.62 | 292.28 | 60.674 | 0.642 | |
| | | 0.6 | 456.71 | 294.59 | 61.449 | 0.645 | |
| | | 0.8 | 458.05 | 295.60 | 62.019 | 0.645 | |
| $TiO_2$ | | 1 | 459.40 | 295.85 | 62.440 | 0.644 | |
| PS (V-rich)$^b$ | 773 | 0.19 | 455.4 | 288.80 | 59.894 | 0.634 | 328 |
| | 673 | 0.14 | 455.4 | 287.85 | 59.697 | 0.632 | 333 |
| | 573 | 0.095 | 455.4 | 286.80 | 59.479 | 0.630 | 322 |
| | 473 | 0.096 | 455.3 | 286.95 | 59.484 | 0.630 | |
| PS (Ti-rich)$^b$ | 773 | 0.46 | 455.6 | 293.80 | 60.984 | 0.645 | |
| | 673 | 0.54 | 455.4 | 295.40 | 61.263 | 0.649 | |
| | 573 | 0.58 | 455.5 | 295.90 | 61.393 | 0.650 | |
| | 473 | 0.61 | 455.7 | 296.00 | 61.468 | 0.650 | |

$^a$obtained by rapid quenching from 1173 K into ice water
$^b$obtained by annealing a solid solution with $x(ss) = 0.4$ at $T_a$ for 12 h
$^c$monoclinic ($P2_1/c$), $a = 575.62$ pm, $b = 452.95$ pm, $c = 537.77$ pm, $\beta = 122.65°$
$^d$monoclinic ($P2_1/c$), $a = 579.09$ pm, $b = 453.17$ pm, $c = 537.50$ pm, $\beta = 122.47°$
$^e$monoclinic ($C2/m$), $a = 908.49$ pm, $b = 580.07$ pm, $c = 454.56$ pm, $\beta = 90.53°$

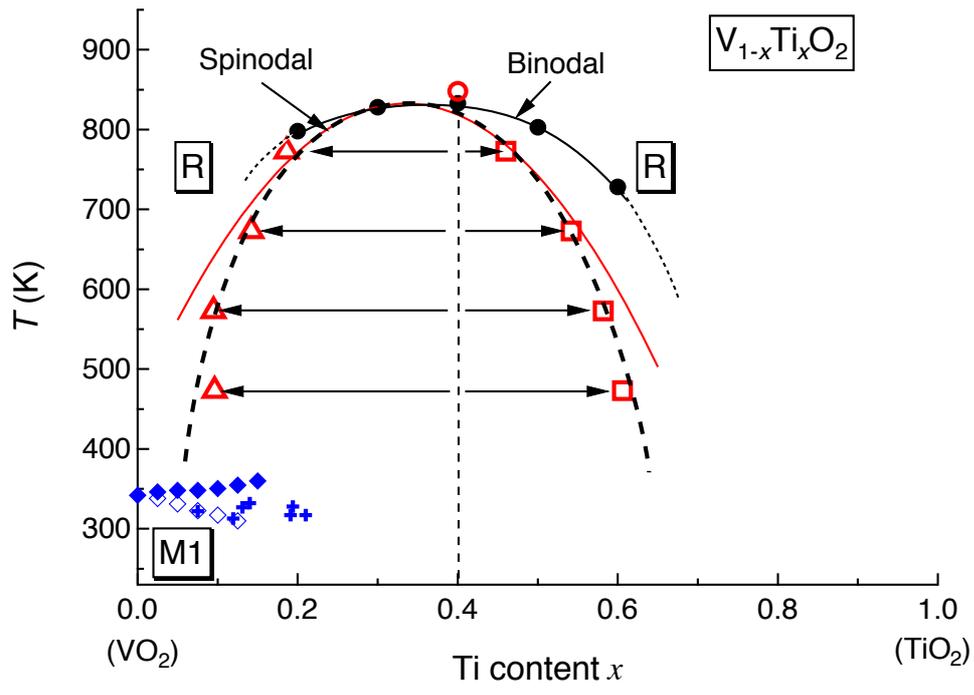